\documentclass[twocolumn,showpacs,showkeys,preprintnumbers,amsmath,amssymb]{revtex4-1}

\usepackage{graphicx}
\usepackage{dcolumn}
\usepackage{bm}

\usepackage{hyperref}
\usepackage[T1]{fontenc} 
\usepackage{color}

\begin{document}

\preprint{Submitted to: JOURNAL OF SUPERCONDUCTIVITY AND NOVEL MAGNETISM}

\title{Metastability and phase separation in a simple model of\\ a~superconductor with extremely short coherence length}%

\author{Konrad Kapcia}%
    \email{e-mail: konrad.kapcia@amu.edu.pl}
\affiliation{Electron States of Solids Division, Faculty of Physics, Adam Mickiewicz University in Pozna\'n, Umultowska 85, PL-61-614 Pozna\'n, Poland, EU
}%


\date{November 14, 2013}

\begin{abstract}
We present studies of the atomic limit of the extended Hubbard model with pair hopping for arbitrary electron density and arbitrary chemical potential. The Hamiltonian consists of (i) the effective on-site interaction $U$ and (ii) the  intersite charge exchange term $I$, determining the hopping of electron pairs between nearest-neighbour sites. In the analysis of
the phase diagrams and thermodynamic properties of
this model we treat the intersite interactions within the mean-field approximation. In this report we focus on metastable phases and determine their ranges of occurrence. Our investigations in the absence of the external magnetic field show that the system analysed exhibits tricritical behaviour. Two metastable phases (superconducting and nonordered) can exist inside the regions of the phase separated state stability and a first-order transition occurs between these metastable phases.
\end{abstract}

\maketitle

\section{Introduction}\label{intro}

The superconductivity (SS) with very short coherence length and the phase separation (PS) phenomenon involving SS states are very current topics (for a review see \cite{MRR1990,AAS2010,DHM2001,KRM2012,KR2013} and references therein). It is worthwhile to mention that metastable and unstable states have been found in many physical systems experimentally and theoretically.

In our work we will study a model which directly pertains to that problem.
The effective Hamiltonian considered has the following form:

\begin{eqnarray}\label{row:ham1}
\hat{H} & = & U\sum_{i}{\hat{n}_{i\uparrow}\hat{n}_{i\downarrow}}- 2I\sum_{\langle i,j\rangle}{\hat{\rho}_i^+\hat{\rho}_j^-} \nonumber\\
& - & \mu\sum_i\hat{n}_i - B\sum_i{\hat{s}^z_i},
\end{eqnarray}
where  \mbox{$\hat{n}_{i}=\sum_{\sigma}{\hat{n}_{i\sigma}}$}, \mbox{$\hat{n}_{i\sigma}=\hat{c}^{+}_{i\sigma}\hat{c}_{i\sigma}$}, \mbox{$\hat{\rho}^+_i=(\hat{\rho}^-_i)^\dag=\hat{c}^+_{i\uparrow}\hat{c}^+_{i\downarrow}$}.
$B=g\mu_BH_z$ is external magnetic field and $\hat{s}^z_i=(1/2)(\hat{n}_{i\uparrow}-\hat{n}_{i\downarrow})$ is $z$-component of the total spin at $i$ site. \mbox{$\sum_{\langle i,j\rangle}$} indicates the sum over nearest-neighbour sites $i$ and $j$ independently.
$\hat{c}^{+}_{i\sigma}$ ($\hat{c}_{i\sigma}$) denotes the creation (annihilation) operator of an electron with spin \mbox{$\sigma=\uparrow,\downarrow$} at the site $i$,
which satisfies canonical anticommutation relations:
\begin{equation*}
\{ \hat{c}_{i\sigma}, \hat{c}^+_{j\sigma'}\}  =  \delta_{ij}\delta_{\sigma\sigma'}, \quad
\{ \hat{c}_{i\sigma}, \hat{c}_{j\sigma'}\}  =  \{ \hat{c}^+_{i\sigma}, \hat{c}^+_{j\sigma'}\} = 0,
\end{equation*}
where $\delta_{ij}$ is the Kronecker delta.
$\mu$ is the chemical potential, connected with the concentration
of electrons by the formula:
\begin{equation*}
n = \frac{1}{N}\sum_{i}{\left\langle \hat{n}_{i} \right\rangle},
\end{equation*}
with \mbox{$0\leq n \leq 2$} and $N$ is the total number of lattice sites. \mbox{$I_0=zI$}, where $z$ is a~number of the nearest-neighbour sites and $\langle\hat{A}\rangle$ indicates the average value of the operator $\hat{A}$ in the grand canonical ensemble.

Model (\ref{row:ham1}) exhibits (in the absence of the field conjugated with the superconducting (SS) order parameter $\Delta=\frac{1}{N}\sum_i \langle \hat{\rho}^-\rangle$) a symmetry between \mbox{$I>0$} ($s$-pairing) and \mbox{$I<0$} ($\eta$-pairing, $\eta$S, \mbox{$\Delta_{\eta S} = \frac{1}{N}\sum_i{\exp{(i\vec{Q}\cdot\vec{R}_i)}\langle \hat{\rho}^-_i\rangle} $}, $\vec{Q}$ being half of the smallest reciprocal lattice vector) cases.
Thus, we restrict ourselves  to the \mbox{$I>0$} case only. In the presence of finite single electron hopping \mbox{$t_{ij} \neq 0$} the symmetry is broken in the general case \cite{HD1993,RB1999,N2,N3,N4}.

Model (\ref{row:ham1}) has been intensively analysed for \mbox{$B=0$} \cite{B1973,HB1977,WA1987,RP1993,R1994,KRM2012} as well as for \mbox{$B\neq0$} \cite{RP1993,KR2013} (in particular, in the context of the phase separation \cite{KRM2012,KR2013}).
In the analysis we have adopted a variational approach (VA), which treats the on-site interaction term
($U$) exactly and the intersite interaction ($I$) within the mean-field approximation (MFA). One obtains two
equations for $n$ and $\Delta$, which are solved self-consistently. Explicit forms of equations for the energy and other
thermodynamical properties are derived in Refs. \cite{KRM2012,KR2013,RP1993}. Condition \mbox{$\Delta\neq 0$} is in the superconducting (SS) phase, whereas in the
nonordered (NO) phase \mbox{$\Delta = 0$}. For fixed $n$, the model can exhibit also the phase separation (PS) which is a state with two coexisting domains (SS and NO) with different electron concentration, $n_-$ and $n_+$. The free energy of the PS state can be derived in standard way, using Maxwell's construction (e.g.~\cite{KRM2012,KR2013,KR2011,KR2011a,B2004}). It is important to find all homogeneous solutions at which grand canonical potential $\omega$ (free energy $f$) has the local minimum with respect to $\Delta$ if system is considered for fixed $\mu$ (or $n$).

We say that the solution (of the set of two self-consistent equations for $n$ and $\Delta$) corresponds to a~metastable phase if it gives a~(local) minimum of $\omega$ (or $f$) with respect to $\Delta$ and the stability condition \mbox{$\partial \mu/\partial n > 0$} (system with fixed $n$) is fulfilled.
Otherwise, we say that the phase is unstable. A~stable (homogeneous) phase is a~metastable phase with the lowest free energy (among  all metastable phases and phase separated states).

In the paper we have used the following convention. A~second- (first-)order transition is a~transition between homogeneous phases  with a~(dis-)continuous change of the order parameter at the transition temperature. A~transition between homogeneous phase and PS state is symbolically named as a~``third-order'' transition \cite{KRM2012,KR2013}. At this transition a~size of one domain in the PS state decreases continuously to zero at the~transition temperature.  We have also distinguished a~first-order transition between metastable phases.

The phase diagrams obtained are symmetric with respect to half-filling because of the particle-hole symmetry of the hamiltonian (\ref{row:ham1}) \cite{MRR1990,KRM2012,KR2013}, so the diagrams will be presented only in the range \mbox{$\bar{\mu}=\mu-U/2 \leq 0$} and \mbox{$0\leq n\leq 1$}.

In present report we will focus on the possibility of the metastable phases occurrence on the phase diagrams of model (\ref{row:ham1}) in the absence of magnetic field (\mbox{$B=0$}). The effects of \mbox{$B\neq 0$} are rather similar to those of \mbox{$U>0$} \cite{KRM2012,KR2013,RP1993} and we leave deeper analysis of the \mbox{$B\neq0$} case to future publications.

\section{Numerical results and discussion (\mbox{$B=0$})}

The overall behaviour of the system has been shown in \cite{RP1993,KRM2012,KR2013}. The model considered exhibits interesting multicritical behaviour including tricritical points.

\begin{figure}
    \centering
    \includegraphics[width=0.49\textwidth]{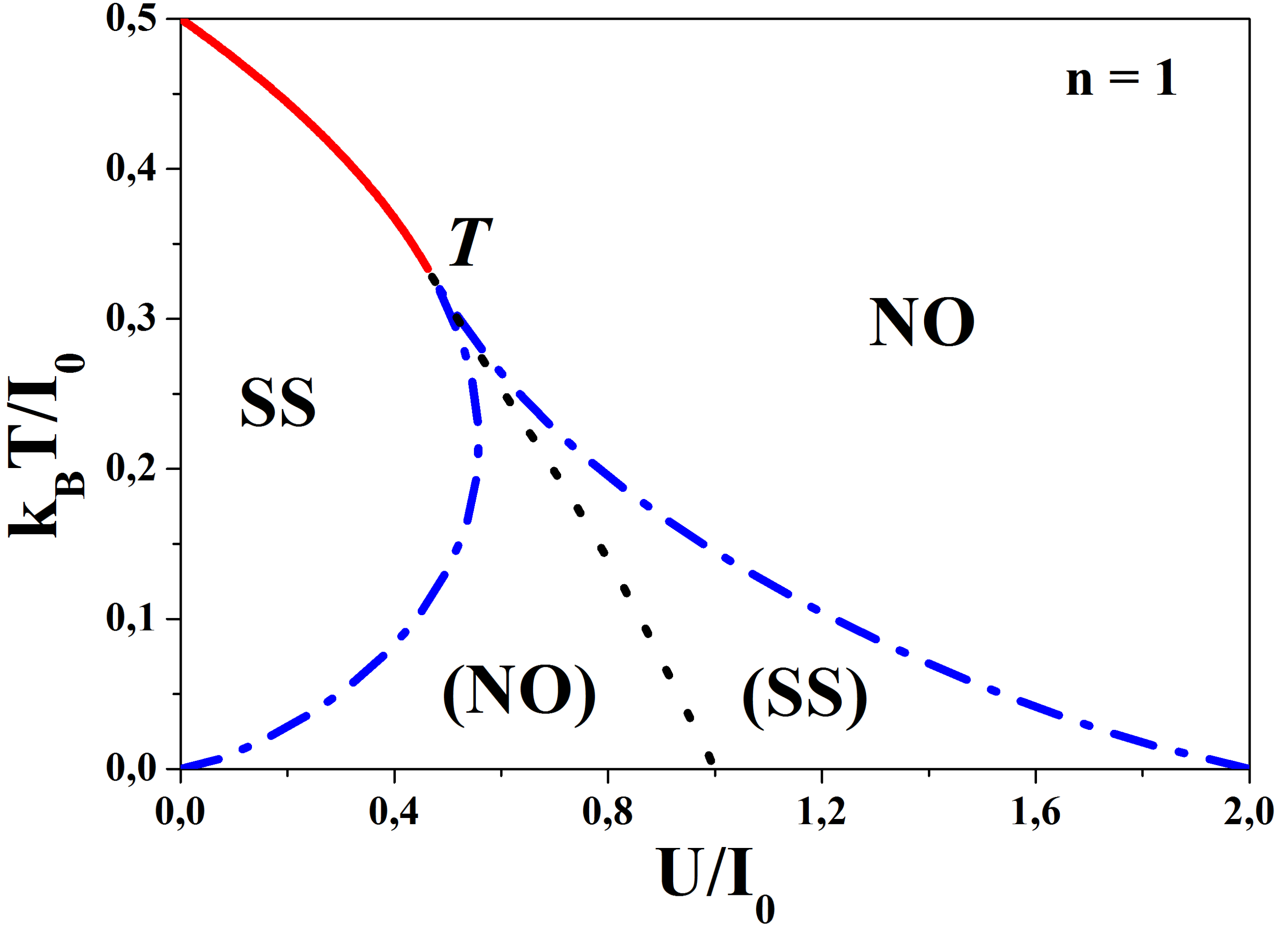}
    \caption{$k_BT/I_0$~vs.~$U/I_0$ phase diagram for \mbox{$n=1$} (\mbox{$I_0=zI$}). Dotted and solid lines denote first-
and second-order transitions between stable phases. Dashed-dotted lines denote the boundaries of metastable phase occurrence
(names of metastable phases in brackets). $T$ denotes tricritical point.}
    \label{fig:diaghalf}
\end{figure}

In the range \mbox{$2<U/I_0<+\infty$}, only the NO phase is stable at any \mbox{$T\geq0$}.
For  on-site attraction \mbox{$U/I_0<0$}, (\textit{``local pair'' limit}) only the second order  \mbox{SS--NO} transitions between homogeneous phases occur with increasing temperature. The transition  temperature is maximal for \mbox{$U\rightarrow - \infty$}, \mbox{$\bar{\mu}=0$} (\mbox{$n=1$}) and it decreases monotonically with increasing $U/I_0$ and $|\bar{\mu}|/I_0=|n-1|$.

The most interesting is the range \mbox{$0<U/I_0<2$}. In this range there is smooth crossover into the \textit{``pair breaking'' limit} and the SS--NO transition can also be of a~first order (for fixed $\bar{\mu}$) and the system exhibits phase separation (for fixed $n$). The metastable phases exist in several definite ranges of model parameters as it will be discussed below.

One should stress that metastable phases can occur only at \mbox{$T>0$}. At \mbox{$T=0$} one phase (state) can be stable only.
For \mbox{$T=0$} the discontinuous SS--NO transition occurs at \mbox{$U/I_0=(\bar{\mu}/I_0)^2+1$} (for fixed \mbox{$|\bar{\mu}|/I_0<1$}) whereas the continuous SS--NO transition occurs at \mbox{$|\bar{\mu}|/I_0=1$} and \mbox{$U/I_0<2$}.
The PS state stability region is determined by conditions: \mbox{$U/I_0\leq2$} and \mbox{$|n-1|^2\leq U/I_0-1$} (\mbox{$n\neq1$}).
At \mbox{$n=1$} (\mbox{$\bar{\mu}=0$}) the discontinuous SS--NO transition occur for \mbox{$U/I_0=1$}.
The extension (to the ground state) of the end of the first order transition line between metastable phases (SS and NO) is located at \mbox{$U/I_0=1+|1-n|$} (for fixed $n$).
The boundaries for the regions of the metastability of homogeneous phases at \mbox{$T>0$} near the ground state are: for the NO phase --- \mbox{$U/I_0=2|\bar{\mu}|/I_0$} and \mbox{$|\bar{\mu}|/I_0<1$} (\mbox{$U/I_0= 2|n-1|$}, any $n$); for the  SS phase --- \mbox{$U/I_0=2$} and \mbox{$|\bar{\mu}|/I_0<1$} (any $n$).
Notice that for both homogeneous phases the condition \mbox{$\partial \mu/\partial n \geq 0$}  is fulfilled at \mbox{$T=0$} (in particular in the ranges of the PS state occurrence) -- cf. Sec.~3 of \cite{KRM2012}.
Let us point out that for \mbox{$T=0$} the discontinuous transition between two NO phases with $|n-1|=1$ and $n=1$ (Mott state) occurs at \mbox{$U/I_0=2|\bar{\mu}|/I_0$} and \mbox{$|\bar{\mu}|/I_0>1$}, but it does not exist for any \mbox{$T>0$}.

\begin{figure*}
    \centering
    \includegraphics[width=1\textwidth]{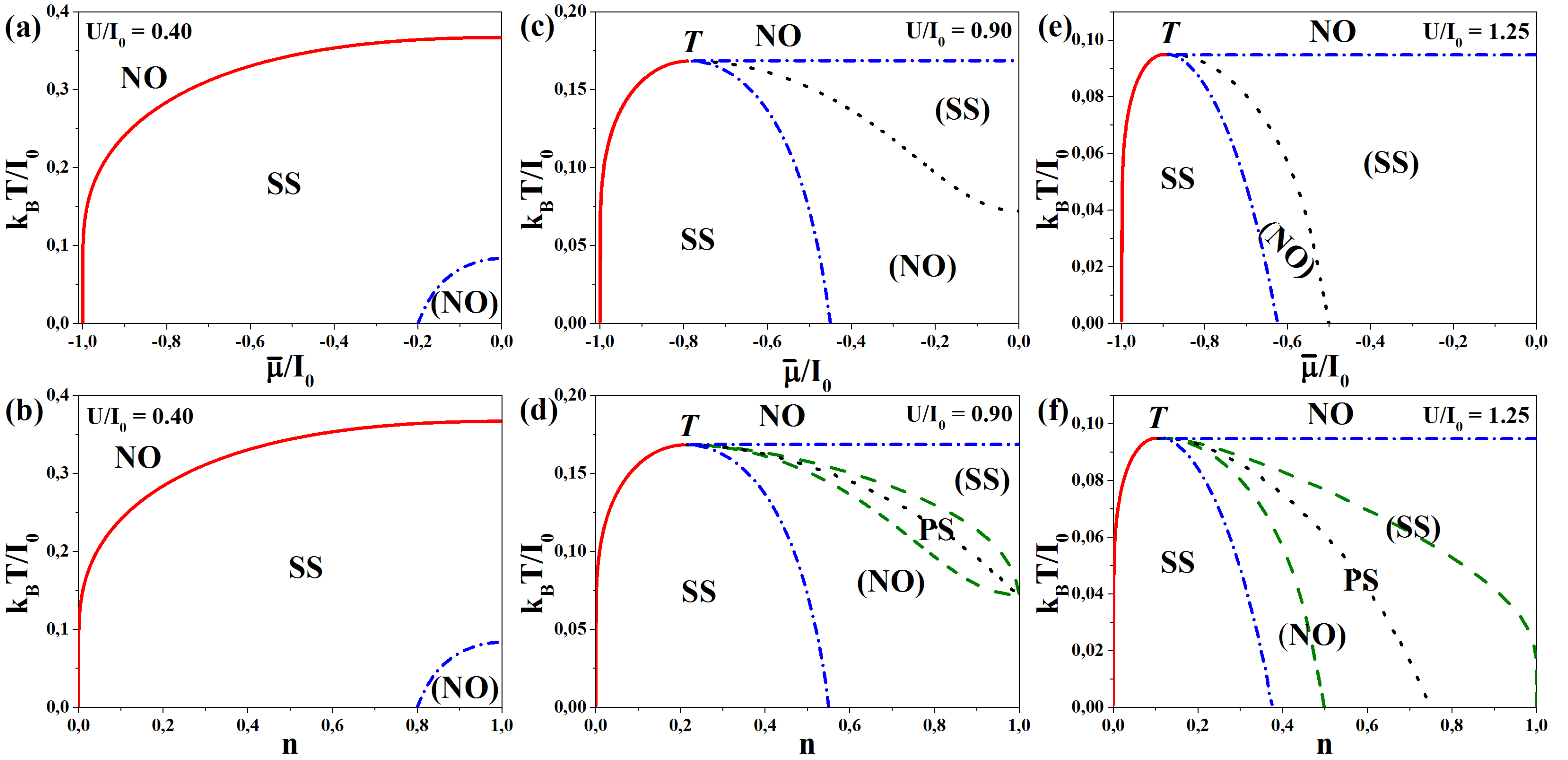}
    \caption{$k_BT/I_0$~vs.~$\bar{\mu}/I_0$ phase diagrams (upper row) and corresponding $k_BT/I_0$~vs.~$n$ diagrams (lower row)  for \mbox{$U/I_0=0.4,\ 0.9,\ 1.25$} (as labelled). Dotted, solid and dashed lines indicate first-order, second-order and ``third-order'' boundaries, respectively. Dashed-dotted lines indicate the boundaries of metastable phase occurrence (names of metastable phases in brackets). $T$ denotes tricritical point.}
    \label{fig:diagmiandn}
\end{figure*}

\subsection{The half-filling (\mbox{$\bar{\mu} = 0$}, \mbox{$n=1$})}\label{sec:half}

In Fig.~\ref{fig:diaghalf} we present the phase diagram involving metastable phases for half-filling (\mbox{$\bar{\mu} = 0$}, \mbox{$n=1$}). One can distinguish four ranges of on-site repulsion, in which a~different behaviour can occur:
\begin{itemize}
\item[(i)]{\mbox{$0<U/I_0<\frac{2}{3}\ln 2$} --- the second-order SS--NO transition is present and at low temperatures the NO phase is metastable;}
\item[(ii)]{\mbox{$\frac{2}{3}\ln 2<U/I_0 < 0.557$} --- the first-order SS--NO transition occurs (it takes place in the whole range \mbox{$\frac{2}{3}\ln 2<U/I_0 < 1$}). Above this transition temperature the SS phase is metastable, whereas the NO phase is metastable (close) below the transition temperature and at low temperatures there is another region where the NO phase is metastable;}
\item[(iii)]{\mbox{$0.557<U/I_0 < 1$} ---  there is one region of metastability of the NO phase, which extends from \mbox{$T=0$};}
\item[(iv)]{\mbox{$1<U/I_0 < 2$} --- there is no transitions with increasing temperature and only the NO phase is stable. At sufficiently low temperatures the SS phase is metastable.}
\end{itemize}

Notice that at \mbox{$n=1$} the VA results for model (\ref{row:ham1}) can be simply mapped onto these of the $U$-$W$ model with \mbox{$W>0$} \cite{R1973,MRC1984,MM2008,KKR2010,KR2011,KR2011a,KR2012}. In such a~case the SS phase corresponds to the charge-ordered phase on the phase diagram \cite{KR2012}.
The results from Fig.~\ref{fig:diaghalf} can be also transformed into the $U$-$J$ model \cite{KKR2010a,MKPR2012,MPS2013} for \mbox{$n=1$} by generalized \mbox{$U\leftrightarrow-U$} Shiba's transformation 
\cite{MRR1990,S1972,RMC1981}.
In such case the SS phase corresponds to the magnetic phase with simultaneous change \mbox{$U\rightarrow-U$} on the diagram in Fig.~\ref{fig:diaghalf}.

\subsection{Arbitrary electron concentrations}

In this section we present results for arbitrary concentration $n$ (and arbitrary chemical potential \mbox{$\bar{\mu}=\mu-U/2$}). A few particular phase diagrams are shown in Fig.~\ref{fig:diagmiandn}. Let us  discuss them in the order which corresponds to the ranges of $U/I_0$ mentioned in Sec.~\ref{sec:half}.

(i):~{\mbox{$0<U/I_0<\frac{2}{3}\ln 2$}. The phase diagrams for \mbox{$U/I_0=0.4$} are shown in Figs.~\ref{fig:diagmiandn}(a,b).
The \mbox{SS--NO} transition between (stable) homogeneous phases is a continuous one and its temperature decreases monotonically with increasing $U/I_0$ and $|\bar{\mu}|/I_0=|n-1|$. Moreover, at sufficiently low temperatures, there is a~region (extending from half-filling) of the NO phase metastability.}

\begin{figure}
    \centering
    \includegraphics[width=0.49\textwidth]{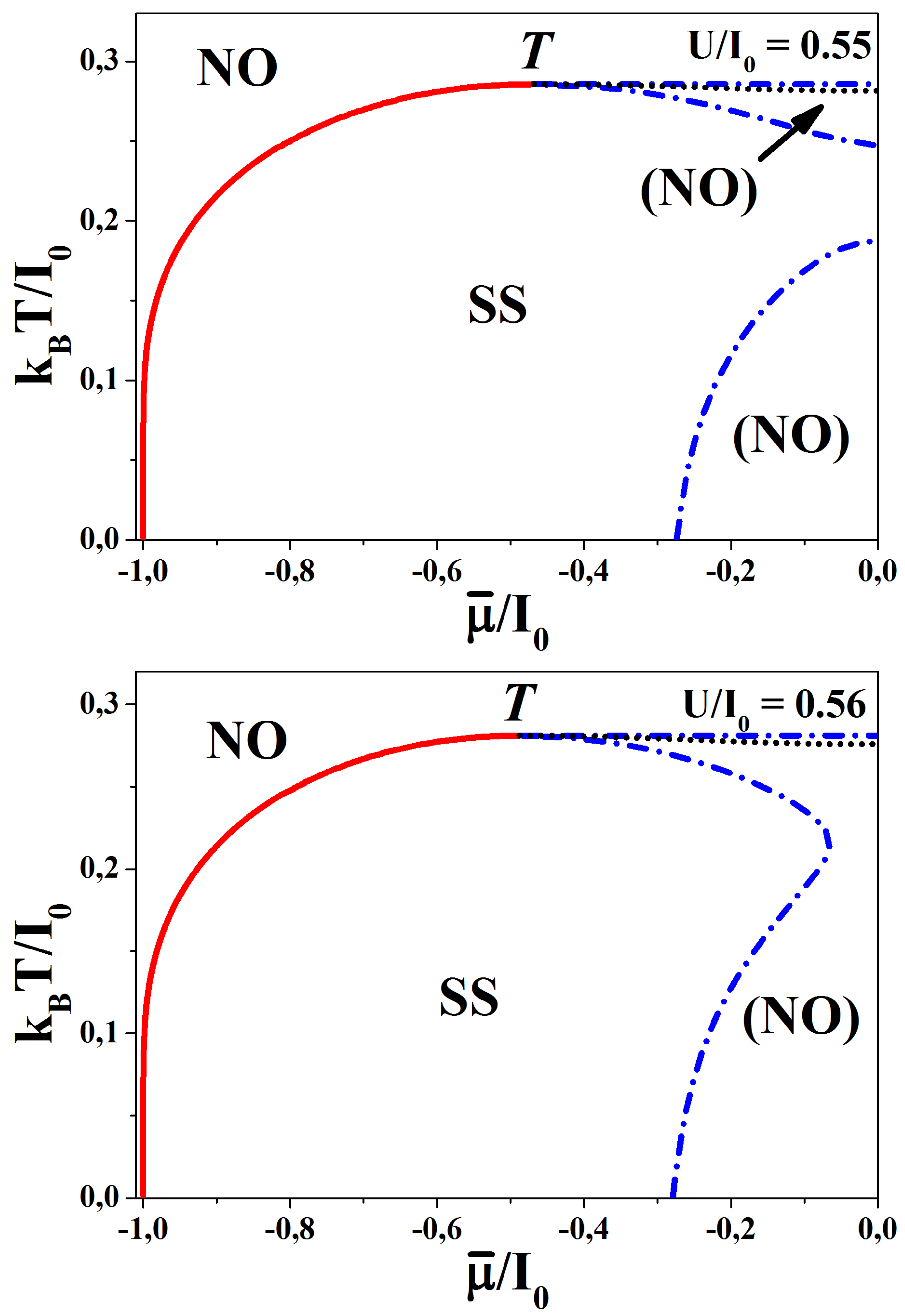}
    \caption{$k_BT/I_0$~vs.~$\bar{\mu}/I_0$ phase diagrams for \mbox{$U/I_0=0.55,\ 0.56$} (as labelled). Above the first-order SS--NO boundary a~narrow region of the SS phase metastability is present (not indicated explicitly). Denotations as in Fig.~\ref{fig:diaghalf}.}
    \label{fig:diagnear}
\end{figure}

(ii)/(iii):~\mbox{$\frac{2}{3}\ln 2<U/I_0 < 1$}. With the increasing of $U/I_0$ in the vicinity of \mbox{$n=1$} (\mbox{$\bar{\mu}=0$}) the SS--NO transition changes its order from second order in the first order and the tricritical point $T$ appears on the phase diagram (cf. Fig.~\ref{fig:diagmiandn}(c,d) for \mbox{$U/I_0=0.9$} and Fig.~\ref{fig:diagnear}). It is quite obvious that in the neighbourhood of the first-order SS--NO transition (for fixed $\bar{\mu}$) the regions of the metastable phases occurrence are present (above the  transition temperature  the SS phase is metastable, whereas below the transition temperature the NO phase is metastable).
The first-order SS--NO transition line (on the diagram for fixed $\bar{\mu}$) splits  into two ``third-order'' lines (on the diagram for fixed $n$) and the PS state is stable at \mbox{$T>0$} in definite range of parameters (between the ``third-order'' lines, for \mbox{$n\neq1$}). In the region of the PS state occurrence (in which the PS state has the lowest energy $f_{PS}$) the first-order transition between two metastable (homogeneous) phases (SS, NO)  exists at \mbox{$T>0$}. Above this line the SS phase has the highest energy (i.e. \mbox{$f_{SS}>f_{NO}>f_{PS}$}), whereas below the line the energy of the NO phase is higher than the energy of the SS phase (\mbox{$f_{NO}>f_{SS}>f_{PS}$}). The line of SS--NO first-order transition between metastable phases ends at \mbox{$n=1$} and \mbox{$T>0$}. One metastable phase (SS or NO) can also exist in the regions of homogeneous phases (NO or SS, respectively) stability for fixed $n$ (where the PS state does not exist), cf. Fig.~\ref{fig:diagmiandn}(d).

The only difference between cases (ii) and (iii) is that for \mbox{$U/I_0<0.557$} the separated region of the NO phase metastability exists also at sufficiently low temperatures. For \mbox{$U/I_0\approx0.557$} that region connects with the NO phase region of metastability at higher temperatures (at half-filling, cf. Fig.~\ref{fig:diaghalf} and Fig.~\ref{fig:diagnear}).

(iv):~\mbox{$1<U/I_0 < 2$}. The exemplary phase diagrams for \mbox{$U/I_0=1.25$} are shown in Figs.~\ref{fig:diagmiandn}(e,f). The line of SS--NO first order transition between stable phases (for fixed $\bar{\mu}$) and metastable phases (for fixed $n$)  ends  at \mbox{$T=0$} and \mbox{$\bar{\mu}<0$} (\mbox{$n<1$}). The region of the PS state stability extends from the ground state. The rest of the discussion is similar to the case (ii)/(iii).

The thermodynamic properties of the model have been analysed in \cite{KRM2012,KR2013}, therefore, we refer the reader to these publications. In particular, the behaviour of thermodynamic parameters in the PS state (as well as in the metastable phases) has been widely discussed in Sec.~5 of \cite{KRM2012}.

\section{Concluding remarks}

The results obtained are important for physics of phase transitions as they involve the investigation of metastable phases. They show that the SS phase metastable boundary is not dependent on $n$ and $\bar{\mu}$ for $|\bar{\mu}|$ ($|1-n|$) smaller than those of $T$-point and that the (meta-)stable solutions for the SS phase can exist only for temperatures lower than those of $T$-point. The SS solution can be  stable or metastable and exists only in regions indicated on phase diagrams. On the  contrary, the NO phase solutions exist at any model parameters and temperature. Outside the regions where the NO phase is (meta-)stable, it is unstable. The first-order boundaries found in \cite{RP1993} correspond to transitions between metastable phases.

Notice  that the behaviour of metastable phases in model (\ref{row:ham1}),  where two metastable phases (SS, NO) can exist in the ranges of the PS state stability, is different than that in model $U$-$W_1$-$W_2$ \cite{KKR2010,KR2011,KR2011a,KR2012} (with \mbox{$W_1>0$} and \mbox{$W_2<0$}), where the metastable phases  cannot exist in the PS occurrence regions at sufficiently low temperatures (at \mbox{$T=0$} for \mbox{$W_2<0$}  \mbox{$\partial \mu/\partial n < 0$} for fixed $n$ in all homogeneous phases) \cite{KR2011,KR2011a,KR2012}.

The on-site $U$ term is the main factor determining the pair binding energy  and the on-site density-density fluctuations in the model \cite{RP1993,KRM2012,RB1999,N5}. Due to rigorous treatment of this term within VA our major conclusions of the paper concerning the behaviour of the model are reliable for arbitrary $U$. Moreover, the MFA treatment of the $I$ term is exact in the limit of infinite dimensions and for $I_{ij}$ of infinite range (\mbox{$I_{ij}=\frac{1}{N}I$} for any $i$, $j$, \mbox{$I>0$}) \cite{KRM2012,RP1993,HB1977} (e.g. effective long-range $I_{ij}$ interaction derived from the coupling between the wide band electrons and local pairs \cite{MRR1990}).

The interesting problem is the competition and interplay between superconductivity and charge orderings (generated by density-density interaction) \cite{R1973,MRC1984,MM2008,KKR2010,KR2011,KR2011a,KR2012} or magnetism \cite{KKR2010a,MKPR2012,MPS2013,KKR2012}. Some preliminary results of such investigations have been presented in \cite{KRM2012,RP1996,PR1997,K2013,K2012,N6}.

\begin{acknowledgements}
The author is indebted to Professor Stanis\l{}aw Robaszkiewicz for very fruitful discussions during this work and careful reading of the manuscript.
The work has been financed by National Science Centre (NCN) as a research project in the years 2011--2013, under Grant No. DEC-2011/01/N/ST3/00413 and as a doctoral scholarship in the years 2013--2014 No. DEC-2013/08/T/ST3/00012.
We thank
the European Commission and the Ministry of Science and Higher Education (Poland)
for the partial financial support from the European Social Fund---Operational Programme ``Human Capital''--POKL.04.01.01-00-133/09-00---``\textit{Proinnowacyjne kszta\l{}cenie, kompetentna kadra, absolwenci przysz\l{}o\'sci}'' as well as the Foundation of Adam Mickiewicz University in Pozna\'n for the support from its scholarship programme.
\end{acknowledgements}


\end{document}